\documentclass[pre,twocolumn,showpacs,preprintnumbers]{revtex4-1}
\usepackage{amsmath,bm,xspace,SIunits,braket,color,graphicx}
\usepackage{epsfig,verbatim}

\begin{document}
\preprint{SAND2016-12720J}
\title{Density-functional calculations of transport properties in the non-degenerate limit and the role of electron-electron scattering}
\date{\today}

\author{Michael P. Desjarlais$^{1}$}
\email{mpdesja@sandia.gov}
\author{Christian R. Scullard$^{2}$}
\author{Lorin X. Benedict$^{2}$}
\author{Heather D. Whitley$^{2}$}
\author{Ronald Redmer$^{3}$}

\affiliation{$^{1}$Sandia National Laboratories,  Albuquerque NM 87185, USA}
\affiliation{$^{2}$Lawrence Livermore National Laboratory, Livermore CA 94550, USA}
\affiliation{$^{3}$Institut f{\" u}r Physik, Universit{\" a}t Rostock, Germany}

\begin{abstract}
We compute electrical and thermal conductivities of hydrogen plasmas in the non-degenerate regime using Kohn-Sham Density Functional Theory (DFT) and an 
application of the Kubo-Greenwood response formula, and demonstrate that for thermal conductivity, the mean-field treatment of the electron-electron (e-e) 
interaction therein is insufficient to reproduce the weak-coupling limit obtained by plasma kinetic theories. An explicit e-e scattering correction to the DFT is posited by 
appealing to Matthiessen's Rule and the results of our computations of conductivities with the quantum Lenard-Balescu (QLB) equation. Further motivation of our 
correction is provided by an argument arising from the Zubarev quantum kinetic theory approach. Significant emphasis is placed on our efforts to produce properly 
converged results for plasma transport using Kohn-Sham DFT, so that an accurate assessment of the importance and efficacy of our e-e scattering corrections to the 
thermal conductivity can be made.\end{abstract}
\maketitle
\section{Introduction}
There has been a rapid increase of publications over the past fifteen years on the computation of electrical and thermal conductivities for warm dense matter 
(i.e., from warm liquids to hot dense plasmas) \cite{Desjarlais02, Mattsson06, Kietzmann08, Recoules09, Hansen11,Lambert11,Holst11,Pozzo11,Hu14,Hu16} 
using Kohn-Sham DFT \cite{Hohenberg64, Kohn65}. In these studies, molecular dynamics (MD) simulations are first performed for classical ions moving in the 
force fields provided by the self-consistently determined electron density within the Born-Oppenheimer approximation. The resulting thermally occupied 
Kohn-Sham states from individual ionic snapshots are then inserted into Kubo-Greenwood \cite{Kubo57,Greenwood58} formulas to calculate the appropriate 
current-current correlation functions. Finally, the results from different uncorrelated snapshots are averaged together and electrical ($\sigma$) and thermal ($\kappa$) 
conductivities are obtained. Because the temperatures are high enough so that many electrons are free to conduct, and thermal electrons move so much 
faster than thermal ions, $\sigma$ and $\kappa$ for such systems are governed entirely by the behavior of the electron currents: the charge current ${\bf j}_{e}$, 
for $\sigma$, and the heat current ${\bf j}_{Q}$, for $\kappa$. The calculations then amount to a determination of the degradation of these currents resulting 
from the interactions of the current-carrying electrons with the rest of the plasma, leading to resistance. 

The advantage of using a DFT-based approach for dense plasmas is that it is unnecessary to decide {\it a priori} which electrons are ``bound" and which are ``free", 
as the degree of localization of a given single-electron state is determined in the course of solving the effective mean-field Schr\"odinger-like equation. However, there 
is also a disadvantage: The electron-electron interaction is treated in a manner in which the electrons are considered as an aggregate, through their total charge 
density, rather than individually. This is in sharp contrast to kinetic theory approaches such as the Boltzmann equation, in which explicit encounters between individual 
particles are considered in the collision terms. With the exception of DFT's use of an exchange-correlation potential (which itself depends only on the total electron 
charge density), the treatment of the e-e interaction is essentially equivalent to that in the Vlasov equation; explicit e-e collisions are absent. 

The classic plasma kinetic theory for $\sigma$ and $\kappa$ is that of Spitzer and H\"arm \cite{Spitzer53}, in which a Fokker-Planck equation is solved to determine 
the steady-state electron velocity distribution resulting from the application of ${\bf E}$ or ${\nabla} T$, from which ${\bf j}_{e}$ and ${\bf j}_{Q}$ are calculated. The 
collisions are treated with Coulomb logarithms \cite{Landau36}, $\log\lambda_{ei}$ and $\log\lambda_{ee}$, which account for screening of the two-body interactions 
in a manner suited to the limit of weak plasma coupling (e.g., small-angle scattering). Yet a more sophisticated kinetic theory approach is that of the Lenard-Balescu 
(LB) equation \cite{Lenard60}, in which the bare Coulomb collisions are dressed by the multicomponent wave vector and frequency dependent dielectric function; 
while resulting in answers identical to that of a Fokker-Planck equation with appropriately chosen Coulomb logarithms for sufficiently weak coupling, LB constitutes a 
predictive theory in which arbitrary distributions and particle species can be considered. Conductivities of quantum plasmas as predicted by QLB are available 
\cite{Williams69,Morales89,Redmer90}, and comparisons between LB predictions for classical plasmas and the results of classical MD have proven very favorable for 
comparable regimes of plasma coupling \cite{Bernu88,Whitley15}. However unlike in the DFT treatment, only free electrons are considered, and therefore the bound 
versus free distinction must be made at the outset when studying real plasmas. 

A method which attempts to combine some of the positive features of the DFT-MD and kinetic theory approaches (though predating the former) is the average-atom 
prescription, exhibited generally in the Ziman resistivity formula \cite{Ziman61}. Here, various means (including DFT) are used to compute the interaction between an 
electron and a representative ion, together with its surrounding screening cloud of other electrons. This interaction is then used in scattering theories of varying 
sophistication to produce the electronic contributions to $\sigma$ and $\kappa$ \cite{Perrot95,Sterne07,Rosznyai08,Faussurier10,Starrett16}, once a statistical 
distribution of ionic positions is assumed. If the treatment of the representative ion is sufficiently detailed, bound and free electrons can be treated on similar footing. 
However the method only treats the electron-ion scattering; as in the Kohn-Sham DFT, e-e scattering is not included. This is not a serious restriction for low-$T$ liquid 
metals (for which the original Ziman work was intended \cite{Ziman61}), since the imposition of electron degeneracy, and the associated Pauli blocking, suppresses 
the effects of e-e scattering on the electron distribution function \cite{caveat_shifted}. But it produces results which are in significant disagreement with, for instance, 
the Spitzer-H\"arm theory \cite{Spitzer53}, particularly for high-$T$ and for low-$Z$ ions, since the effects of the e-e interaction are not outweighed by those of e-i 
\cite{Braginski58}. As such, it is customary to employ separate multiplicative ``Lorentz gas" \cite{Lorentz05} corrections to $\sigma$ and $\kappa$ as determined from 
average-atom theories, which ensure that the final results agree with the weak-coupling limits of plasma kinetic theory \cite{Epperlein86,Sterne07,Rosznyai08}. 
These corrections reduce $\sigma$ and $\kappa$ at high-$T$, relative to their values as predicted by theories in which an e-e collision term is absent, such as in 
average-atom descriptions and in specialized Lenard-Balescu treatments focused primarily on degenerate electrons \cite{Ichimaru85}.

Ziman resistivity formula results 
which are fit to expressions employing Coulomb logarithms, together with $T$-dependent corrections accounting for e-e interaction, form the basis for many 
wide-ranged models for plasma conduction \cite{Lee84,Rinker88,Desjarlais01} used in continuum simulations of astrophysical objects \cite{astro1,astro2, astro3}, 
inertial confinement fusion (ICF) \cite{Hammel10}, and pulsed power applications \cite{ZMatzen}. Indeed, the importance of these applications for dense plasmas has 
fueled several of the DFT-based investigations mentioned at the outset \cite{Recoules09, Hansen11,Lambert11,Holst11,Hu14,Hu16}. As such,  these DFT works 
featured comparisons to some of these models for $\sigma$ and $\kappa$. In Ref.~\cite{Lambert11} for instance, comparisons were made to the high-$T$ limit of the 
Lorenz number (${\kappa}/{\sigma T}$) for hydrogen plasmas, as predicted by Spitzer-H\"arm \cite{Spitzer53}. Though reasonable agreement was found, it has since 
been established that this agreement was spurious, resulting from an incomplete convergence of the DFT calculation of $\kappa$ with respect to the number of 
Kohn-Sham states included in the computation \cite{Desjarlais_private}. While it is certainly reasonable to use DFT-based approaches.
\cite{Hansen11,Lambert11,Holst11,Hu14,Hu16} to attempt to go beyond the many approximations inherent in more conventional plasma descriptions.
\cite{Spitzer53,Williams69,Morales89,Lee84,Rinker88,Desjarlais01}, it is equally important to uncover potential weaknesses in the assumptions underlying current 
implementations of Kohn-Sham DFT for plasma conduction. This in turn requires that these DFT-based predictions of $\sigma$ and $\kappa$ are well-converged.
In this work we use the Kohn-Sham DFT prescription, complete with DFT-based MD and the Kubo-Greenwood approach mentioned above, to predict $\sigma$ 
and $\kappa$ for hydrogen plasmas at sufficiently high-$T$ to make a meaningful comparison to the predictions of the quantum LB equation. Plasma conditions are 
chosen to be $\rho= $ 40~g/cm$^{3}$ and $T$ between 500 eV and 900 eV, to coincide with a previous study \cite{Whitley15} of hydrogen using classical MD and 
statistical potentials \cite{Kelbg63,Dunn67}, where it was demonstrated that the weak-coupling assumptions underlying LB are valid. These conditions offer the added 
advantage that the hydrogen atoms are fully ionized, removing a potential discrepancy between the two approaches.  We show that our DFT prediction for $\sigma$ 
is in excellent agreement with that of LB, while our prediction of $\kappa$ is far too high in this regime. From this, we posit that there are two {\it distinct} contributions 
of the e-e interaction to plasma conduction: 1. A mean-field reshaping of the electron distribution function which is present in the DFT (as well as in any theory 
containing a Vlasov or Hartree term), and 2. A binary e-e scattering piece which is missing in our current implementation of DFT, but which is present in various 
plasma kinetic theories (Fokker-Planck, LB, etc.). We argue that while the first contribution affects both $\sigma$ and $\kappa$, the second contribution only plays a 
role for $\kappa$, due to the inability of binary e-e scattering to degrade the electron {\it charge} current in a system in which the conservation of total electron 
momentum is mandated. By alternately turning off the e-e and e-i collision terms in the quantum LB equation, we demonstrate that an e-e scattering correction to the 
DFT thermal conductivity can be written in the form: $1/\kappa = 1/\kappa_{\rm DFT} + 1/\kappa^{\rm OCP}_{ee}$, where $\kappa^{\rm OCP}_{ee}$ is the thermal 
conductivity of the electron one-component plasma (OCP) as predicted by QLB. We also find that the reshaping contributions to $\sigma$ and $\kappa$ are 
practically identical.  
This general framework is further justified by appealing to arguments derived from a quantum Boltzmann theory using the Zubarev approach 
\cite{Zubarev,Roepke88,Redmer97,Kuhlbrodt00}. Our conclusions extend and amplify those made in a recent work investigating the effect of e-e scattering on the 
electrical conductivity of plasmas \cite{Reinholz15}. 

The remainder of this paper is organized as follows: In Section II, we outline the specific methods we use to produce converged results for the 
conductivities of hydrogen plasmas at high-$T$ using Kohn-Sham DFT. In Section III, we discuss the comparison of these DFT results to those 
of QLB, and construct our e-e scattering correction for $\kappa$; additional motivation for this correction using a quantum Boltzmann approach 
is deferred to the Appendix. We conclude in Section IV. 

\section{The DFT methods}

Density functional molecular dynamics (DFT-MD) simulations, performed with VASP \cite{Kresse93}, are used to generate atomic configurations for 
hydrogen with $\rho = $ 40~g/cm$^{3}$ and $T$ of 500, 700, and 900 eV.  The electronic temperature is established through a Fermi occupation of 
the electronic states.  All calculations are performed in the Local Density Approximation \cite{Kohn65,Ceperley80}. Given the very high densities being 
explored ($r_s = 0.41$ bohr), we employ a bare proton $1/r$ potential for the hydrogen atom.  The plane wave cutoff energy is set to 3800 eV, and the electronic density 
and single-particle wave functions are sampled at a single $\bf k$-point at the $\Gamma$-point (${\bf k}= 0$) in the Brillouin zone corresponding to 
the supercell (see below). The practical limit on our calculations proved to be 256 hydrogen atoms in a periodically-repeated cubic supercell.  
Simulations with more atoms were intractable in combination with the very high temperatures and corresponding need for a very large number of 
bands for the transport properties and the very high plane wave cutoff energy.

In each of these three cases, the electrons are fully ionized from the hydrogen nuclei.  Correspondingly, we calculate the fundamental dimensionless 
plasma parameters for these three cases, namely the ion-ion coupling factor $\Gamma_{ii} \equiv e^2/(k_{B}TR_i) $, where $R_i = (3/(4 \pi n_i))^{1/3}$ 
is the Wigner-Seitz radius for the protons,  and the electron degeneracy $\theta \equiv k_{B}T/E_{\rm Fermi}$ as shown in Table~\ref{plasmanumbers}.  
Note that even for the highest temperature of 900 eV, where $\theta \sim 3$, we expect some residual consequences of electron degeneracy.

\begin{table}
\begin{center}
\begin{tabular}{| c | c | c |}
\hline
$k_{\rm B}T$ (eV)  & $\Gamma_{ii}$ & $\theta $\\ \hline
500 & 0.13 & 1.65\\ \hline
700 & 0.10 & 2.31\\ \hline
900 & 0.07 & 2.97\\ \hline
\end{tabular}
\caption{Dimensionless plasma parameters $\Gamma_{ii}$ and $\theta$ for a fully ionized 40~g/cm$^3$ hydrogen plasma.}
\label{plasmanumbers}
\end{center}
\end{table}

Following the DFT-MD simulations, a set of atomic configurations, well separated in time, are selected for subsequent calculation of the transport 
properties following the treatment in Ref.~\cite{Holst11}.  Achieving convergence on the electrical and thermal conductivities for these high density, 
high temperature systems requires a very large number of bands, from several thousand to well in excess of ten thousand.  A simple one shot 
calculation with such high band numbers is unfeasible owing to poor convergence during the self-consistent determination of the electronic density 
and Kohn-Sham wave functions.  We resort to a stepwise approach building up successively more bands by doing a sequence of calculations with 
increasing band numbers and using prior runs to initialize the simulation.  A consequence of the high band numbers is the need to continue increasing 
the plane wave cutoff energy as there must be plane waves of sufficient energy to represent the highest bands.  The combination of these two 
requirements leads to very poor scaling as the temperature is increased.  

Using the Kubo relation \cite{Kubo57, Greenwood58} for the current-current correlation functions ($\langle {\bf j}_{e}{\bf j}_{e}\rangle$ 
for $L_{11}$, $\langle{\bf j}_{Q}{\bf j}_{Q}\rangle$ for $L_{22}$, etc.), one obtains \cite{Holst11}

\begin{eqnarray}
\label{LDFT}
L_{ij}(\omega) & = & \frac{2\pi (-1)^{i+j}}{3Vm^{2}\omega}\sum_{{\bf k}\nu\mu}  \left( f_{{\bf k}\nu} - f_{{\bf k}\mu} \right)%
\left|\langle {\bf k}\mu |\hat{\bf p}|{\bf k}\nu \rangle\right|^2 \cr
 & \times & \epsilon_{{\bf k} \nu\mu}^{i + j -2} \delta (E_{{\bf k}\mu} - E_{{\bf k}\nu} - \hbar\omega),
\end{eqnarray}
where $i$ and $j$ are labeled by $1$ and $2$ for  charge and heat currents, respectively. $V$ is the system volume, ${\bf k}$ is the electron 
wave vector, $\nu$ and $\mu$ are electron band indices, and $f_{{\bf k}\nu,\mu}$ are the corresponding Fermi occupations.  The $E_{{\bf k}\nu}$ 
are the Kohn-Sham band energies and $\langle {\bf k}\mu |\hat{\bf p}|{\bf k}\nu \rangle$ are the dipole matrix elements.  The Onsager weights are 
given by $\epsilon_{{\bf k}\nu\mu} \equiv \frac{1}{2}(E_{{\bf k}\nu} + E_{{\bf k}\mu}) - h$ where $h$ is the enthalpy per electron. The appearance of 
the wave vector, ${\bf k}$, assumes that we are dealing with a periodic system (supercell, in our case). Though we are ultimately interested in 
DC ($\omega$ = 0) conductivities in this work, we perform computations of the $L_{ij}(\omega)$ for small values of $\omega$ and take the 
limit $\omega\to 0$ , from which we compute $\sigma$ and $\kappa$.

The optical conductivity is given by
\begin{equation}
  \label{eq:electrical-conductivity}
  \sigma(\omega)=e^2{L}_{11},
\end{equation}
with the DC conductivity obtained in the limit $\omega\to 0$.
The convergence of the optical conductivity with respect to the number of bands in the system is readily checked through
the sum rule~\cite{sumrule}
\begin{equation}\label{eqn:sumrule}
S = \frac{2m V}{\pi e^2 N_e}\int_0^\infty \sigma(\omega)d\omega = 1.
\end{equation}
Likewise, the thermal conductivity is obtained from
\begin{equation}
  \label{eq:thermal-conductivity}
  \kappa=\frac{1}{T} \biggl( {L}_{22} - \frac{{L}_{12}\times{L}_{21}}{{L}_{11}} \biggr).
\end{equation}

There are two fundamental challenges in evaluating $\sigma$ and $\kappa$ using the expressions above: 1. The  $\omega \to 0$ limit can be 
problematic, because a finite-sized cell of electrons always possesses a nonzero minimum energy gap (and hence a nonzero minimum value 
of $E_{{\bf k}\mu} - E_{{\bf k}\nu}$) even though an infinite collection of electrons at sufficient density generally does not. Thus, it is necessary 
to determine the DC limit by fitting the $\sigma(\omega)$ and $\kappa(\omega)$ results to $\omega$-dependent forms which have the correct 
behavior for an infinite system while extrapolating the simulations to infinite size.  For the systems considered here, where the electronic density 
of states is free-electron like, and the optical conductivity is well described with the Drude formula, this extrapolation of  $\omega \to 0$ is 
straightforward.  2. For the high temperature plasmas of our interest here, it is necessary to use a {\it very} large number of bands, $(\mu,\nu)$, 
in order to saturate the values of the $L_{ij}$. This is especially true for $L_{22}$ needed for $\kappa$, since the larger power of factors involving 
the single-particle energies more heavily weighs high-energy states. For this, we find it necessary to extrapolate our conductivities to an infinite 
number of bands (or, equivalently, an infinite maximum eigenvalue) by performing a series of  calculations using an increasing number of bands 
for each density and temperature condition we study. 
\begin{figure}[!h]
  \centering
  \includegraphics[width=0.47 \textwidth]{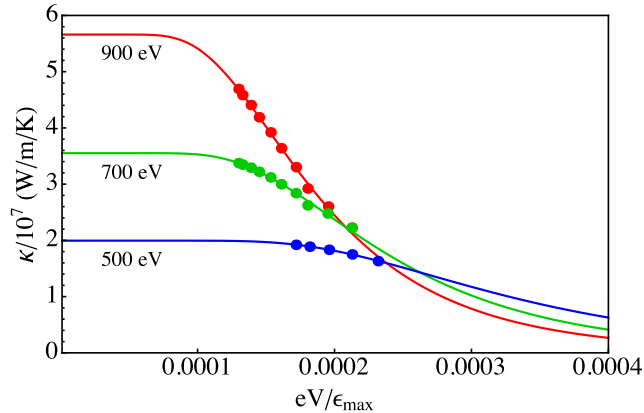}
  \caption{(Color online) $\epsilon_{\rm max} \to \infty$ scaling fit to the thermal conductivity for $T = $ 500, 700, and 900 eV and $\rho = $ 40~g/cm$^3$.}
  \label{fig:kscaling}
\end{figure}
We find that the assumption of a simple power law behavior for the dipole matrix elements describes the asymptotic scaling of the thermal 
conductivity calculations with the maximum eigenvalue very well. 
Noting that $L_{22} \sim E^2$ and representing the dipole matrix elements in the limit $\omega \to 0$ by $E^\gamma$ we fit a series of 
calculations with increasing maximum eigenvalue $\epsilon_{\rm max}$ to the following functional form
\begin{equation}
\label{toinfinity}
\kappa(\epsilon_{\rm max})= \kappa_\infty \frac{\int_{-\infty}^{\epsilon_{\rm max}} E^2 E^\gamma \frac{\partial f}{\partial E} dE}{\int_{-\infty}^{\infty}%
E^2 E^\gamma \frac{\partial f}{\partial E}dE},
\end{equation}
where $f$ gives the Fermi occupations for the temperature and Fermi energy of the system in question.  The values of $\gamma$ and $\kappa_\infty$ 
are then chosen for best fit to the series of calculations for each  $\epsilon_{\rm max}$ at a given temperature.   The results of these fits for the thermal 
conductivity are displayed in Fig.~\ref{fig:kscaling}.   The assumed functional form captures the behavior of the calculated thermal conductivity very 
well in the limit of high $\epsilon_{\rm max}$, giving us high confidence in the resulting value of $\kappa_\infty$.  The best fit values of $\gamma$ 
varied little between the three cases, ranging from 3.3 at 900 eV to 3.4 at 500 eV.  

\begin{figure}[!h]
  \centering
  \includegraphics[width=0.47 \textwidth]{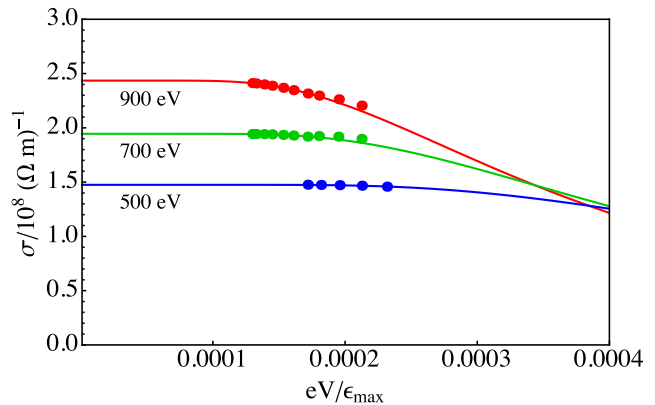}
  \caption{(Color online) $\epsilon_{\rm max} \to \infty$ scaling fit to the electrical conductivity for $T = $ 500, 700, and 
  900 eV and $\rho = $ 40~g/cm$^3$.}
  \label{fig:Sscaling}
\end{figure}

We show the results of the same procedure (as in Eq.~\ref{toinfinity}, but with $E^2 \to 1$) applied to the electrical conductivity in Fig.~\ref{fig:Sscaling}.   
Note the significantly more rapid convergence of the electrical conductivity with increasing $\epsilon_{\rm max}$.     It is important to note 
that even under conditions in which the sum rule (\ref{eqn:sumrule}) on $\sigma(\omega)$ is satisfied to a high degree, the calculation 
of $\kappa$ could still be substantially in error.  For example, the sum rules for the 900 eV case range from 93.9\% at the lowest 
$\epsilon_{\rm max}$ to 98.5\% at the highest $\epsilon_{\rm max}$.

Given our added confidence in these extrapolated predictions of $\sigma$ and (especially) $\kappa$ in these conditions, relative to 
earlier predictions \cite{Lambert11}, we are now in a position to compare them to the results of other approaches.

\section{Comparisons between Kohn-Sham DFT and Lenard-Balescu}

In the moderate-to-weak plasma coupling regime of our interest in this work, we know of no highly constraining experimental results for $\sigma$ 
or $\kappa$ for hydrogen plasmas. Therefore, we can only compare our extrapolated Kohn-Sham DFT results to the predictions from other theories. 
Fortunately, there is an {\it ab initio} plasma kinetic theory which should provide very accurate estimates in this particular regime: 
Lenard-Balescu theory \cite{Lenard60}. A recent work by some of us \cite{Whitley15} demonstrated that {\it classical} LB theory reproduces MD 
computations of $\sigma$ and $\kappa$ for a semiclassical model of hydrogen (in which statistical two-body interaction potentials were 
used \cite{Kelbg63,Dunn67}) for the very same conditions we study here. The primary approximations in LB theory 
pertain to the neglect of large-angle scattering and a specific treatment of density fluctuations as modeled by the Random Phase 
Approximation (RPA) \cite{Ichimaru92}. Since these approximations play very similar roles in both classical and quantum variants of the theory, 
we take the excellent agreement displayed in Ref.~\cite{Whitley15} as a strong indication of the validity of quantum-LB here \cite{caveat_indeed}.

The mathematical and numerical prescription we use to generate quantum-LB predictions of $\sigma$ and $\kappa$ is outlined briefly in Section 2.2 of 
Ref.~\cite{Whitley15}, and the results we show here are in fact identical to those plotted as the thick dark blue lines in Figs.~1 and 2 of that work. 
We note that these predictions are extremely close to those of Williams and DeWitt \cite{Williams69} for $\sigma$ and $\kappa$ derived from 
the quantum-LB equation, though with the minor caveat we mention in Ref.~\cite{caveat_WD}.

For the purposes of the discussion which follows, it is important to understand that the LB kinetic equation for hydrogen possesses two collision 
terms, $C_{ei}$ and $C_{ee}$, each of which involve \cite{Lenard60,Williams69,Morales89,Whitley15}: 1. The Fourier transforms of the bare two-body 
interactions; 2. Occupation factors, $f$, evaluated at the momenta of the colliding particles; and 3. The two-species dynamical RPA dielectric function, 
$\epsilon(q,\omega)$, evaluated at frequencies involving the center-of-mass energies of the colliding particles. In practice, the effects of quantum 
diffraction manifest through the occupation factors, while the effects of screening arise through the dielectric function.  

\begin{figure}[!h]
\begin{center}
\includegraphics[width=0.47 \textwidth]{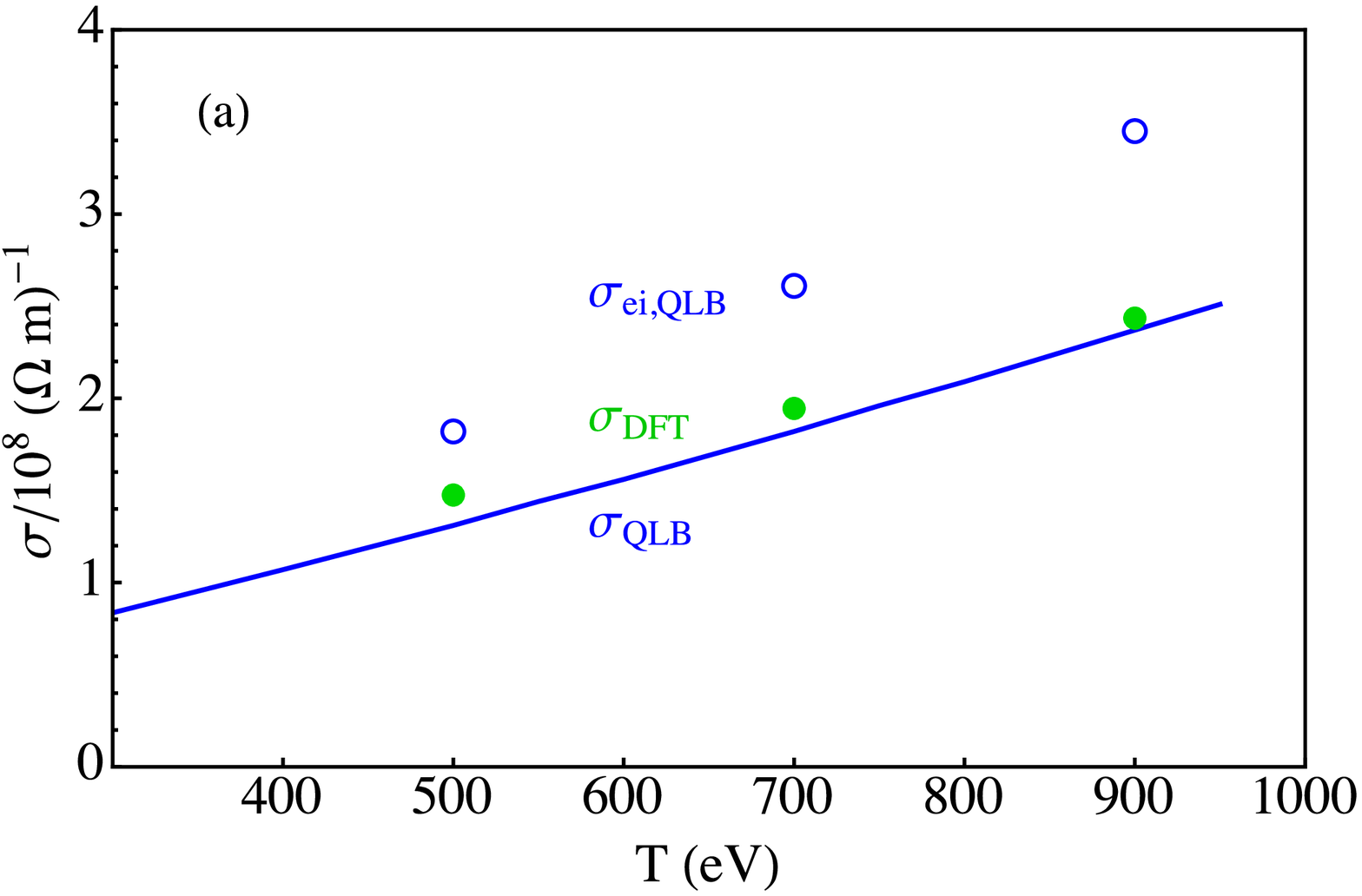}
\includegraphics[width=0.47 \textwidth]{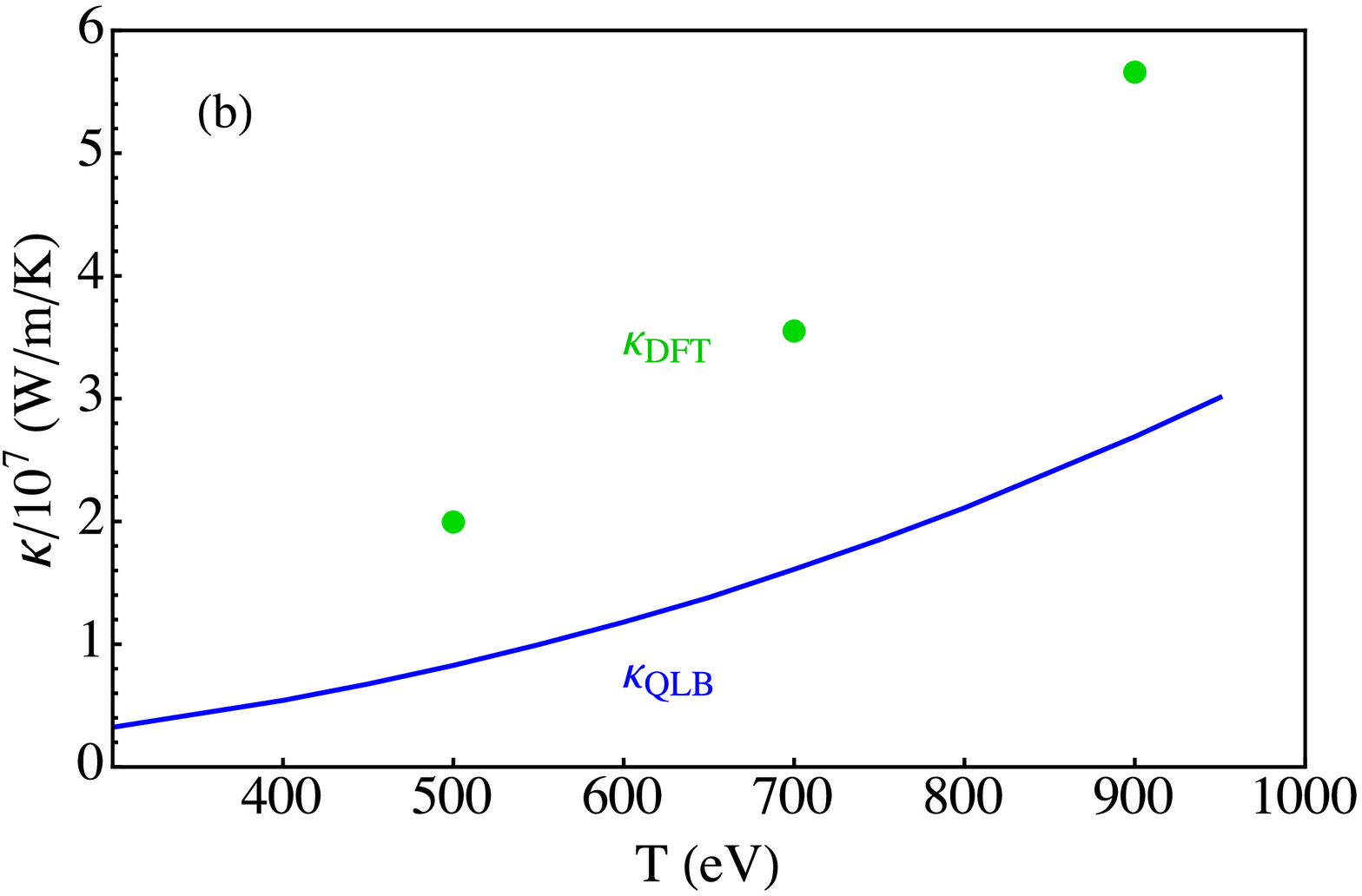}
\caption{(Color online) (a) Electrical conductivity of hydrogen at $\rho=$ 40 g/cm$^{3}$ as computed by Kohn-Sham DFT, extrapolated to an infinite number of 
single-particle states, $\sigma_{\rm DFT}$ (solid green circles); as computed by the QLB equation using the prescription outlined in Ref.~\cite{Whitley15} 
(blue curve); the electrical conductivity in the absence of e-e collisions from the QLB calculations, $\sigma_{ei}$ (open blue circles). (b) Thermal conductivity 
of hydrogen at $\rho=$ 40 g/cm$^{3}$ extrapolated to an infinite number of single-particle states, $\kappa_{\rm DFT}$ (solid green circles); and as computed 
by the QLB equation using the prescription outlined in Ref.~\cite{Whitley15} (blue curve).}
\label{figsigma}
\end{center}
\end{figure}

Fig.~\ref{figsigma}a shows our extrapolated DFT results (solid green circles) for the electrical conductivities of hydrogen plasmas along the 
$\rho= $ 40~g/cm$^{3}$ isochore, as a function of temperature. The general increase of $\sigma$ with $T$ is expected from all theories 
\cite{Spitzer53,Williams69,Ichimaru85,Sterne07}, provided that $T > T_{\rm Fermi}$, as is the case here. Furthermore, the precise magnitude 
is very much in line with our calculation of $\sigma$ using QLB theory \cite{Lenard60,Whitley15}, shown as the blue curve \cite{caveat_WD}. 
The slightly lower $\sigma$ values from QLB can be attributed to our neglect of electron degeneracy in the QLB calculation, 
given that $k_{\rm B}T_{\rm Fermi}$ is as high as 303 eV at this density~\cite{degeneracy}. Though quantum diffraction is accounted for in 
our implementation of QLB, Pauli blocking is not, as the collision terms we use do not possess the proper $1 - f$ factors needed to account 
for Pauli exclusion~\cite{Whitley15}. Nevertheless, the good agreement shown here establishes that upon extrapolation, our Kohn-Sham 
Kubo-Greenwood calculation of $\sigma$ accounts for the bulk of the physics also included in the Lenard-Balescu treatment. 
This physics involves not only scattering of the conducting electrons off the spatially distributed ions dressed by their individual dynamic 
screening clouds, but also the contribution of the e-e interaction in determining the precise shape of the steady-state electron distribution, 
$f(v)$ \cite{caveat_v}. The tendency of the e-e collision term within kinetic theory to reshape the distribution at high-$T$ is well-known in 
the literature; if the simple assumption of the shifted equilibrium distribution \cite{caveat_shifted} is made, $\sigma$ is too low by a factor 
of 1.97 \cite{Ichimaru85}. Indeed, this fact necessitates the application of correction factors when theories which make this assumption are 
used \cite{Sterne07,Rosznyai08,Epperlein86}. Though explicit e-e {\it collisions} are not included in the DFT, it is clear from this comparison that 
the mean-field Hartree (or Vlasov) term is allowing for the proper reshaping of the distribution upon the application of a weak uniform ${\bf E}$-field, 
since the precise magnitude of $\sigma$ is very sensitive to the shape of $f(v)$ \cite{Spitzer53,Ichimaru85}.  
The electrical conductivity in the absence of this proper reshaping contribution of electron-electron collisions, $\sigma_{ei}$, 
as calculated with the QLB equations, is shown with open circles in Fig.~\ref{figsigma}a for comparison
(see the discussion below for our precise definition of $\sigma_{ei}$).  

Figure \ref{figsigma}b shows the corresponding comparison for thermal conductivity. Here, the extrapolated DFT values are higher than those of QLB 
by around a factor of two in this regime. Prior to the realization that this extrapolation was necessary here, the (under-converged) DFT predictions of 
$\kappa$ would have been in better agreement with the QLB results \cite{Lambert11}. As for $\sigma$, $\kappa$ is also known to be affected by the 
e-e interaction within a plasma kinetic theory framework \cite{Spitzer53,Williams69,Ichimaru85}. The correction factor needed to account for its 
effects, relative to a theory in which the low-$T$ shifted equilibrium distribution \cite{caveat_shifted} is assumed, is distinct from that needed for 
$\sigma$. This difference is the combined result of the different forcing terms on the left-hand side of the kinetic equation (corresponding to 
$\nabla T$ rather than ${\bf E}$ \cite{Chapman70}), and the marked differences between the dependences of ${\bf j}_{e}$ and ${\bf j}_{Q}$ on the 
electron velocities. In particular, within a semiclassical framework (see Ref.~\cite{Holst11} and the previous section for more precise expressions 
using Kohn-Sham states), ${\bf j}_{e} \propto \sum_{i} e {\bf v}_{i}$, where $i$ indexes individual electronic states. Note that this is proportional to the 
total electron momentum, $\sum_{i} m{\bf v}_{i}$. We therefore expect individual two-body e-e scatterings to do nothing to degrade ${\bf j}_{e}$, for 
the same reason that these intra-species collisions must leave the total electron momentum unchanged. Because of this, the electron one-
component plasma (OCP) has infinite static electrical conductivity; the application of a constant electric field to a uniform electron gas results in 
resistance-less current. 
In contrast, ${\bf j}_{Q} \propto \sum_{i}(\frac{1}{2}m v_{i}^{2}){\bf v}_{i} \sim \sum_{i}\frac{1}{2}mv_{i}^{3}$, assuming that the potential 
energy contributions to the heat current are negligible in comparison to the kinetic ones, as is the case for the weak coupling conditions studied here 
\cite{Whitley15}. Two-body e-e scatterings can therefore change ${\bf j}_{Q}$ (due to the fact that ${\bf j}_{Q}$ is no longer proportional to a 
conserved quantity) and this results in a finite thermal conductivity for an electron OCP \cite{Chapman70}. It is then reasonable to expect that an 
extra contribution to $\kappa$ from the e-e interaction may result which, in contrast to $\sigma$, depends explicitly on e-e {\it collisions}. This 
beyond-Vlasov/Hartree effect would indeed be absent from the Kohn-Sham DFT prescription we employ here \cite{Reinholz15}.

With these observations in mind, we posit the following relations inspired by Matthiessen's rule \cite{Matthiessen58}, in which these 
two distinct manifestations of e-e interaction, 1. mean-field reshaping of $f(v)$, and 2. binary scattering degradation of ${\bf j}$, are 
added ``in series":
\begin{equation}
\label{eqSsigma}
\frac{1}{\sigma}= \frac{1}{S_{\sigma}\sigma_{ei}} + \frac{1}{\sigma^{\rm OCP}_{ee}}= \frac{1}{S_{\sigma}\sigma_{ei}},
\end{equation}
\begin{equation}
\label{eqSkappa}
\frac{1}{\kappa}= \frac{1}{S_{\kappa}\kappa_{ei}} + \frac{1}{\kappa^{\rm OCP}_{ee}}.
\end{equation}
$\sigma^{\rm OCP}_{ee}$ and $\kappa^{\rm OCP}_{ee}$ are the electrical and thermal conductivities of the electron OCP, which can in principle be 
obtained by scaling down the $C_{ei}$ collision term in a kinetic equation otherwise possessing both $C_{ee}$ and $C_{ei}$ pieces. $\sigma_{ei}$ 
and $\kappa_{ei}$ are the conductivities obtained by turning off the e-e interaction in the precise manner discussed in the following paragraph. The 
second equality in Eq.~\ref{eqSsigma} arises from the fact that $\sigma^{\rm OCP}_{ee}= \infty$, as mentioned above. The factors $S_{\sigma}$ and 
$S_{\kappa}$ are the reshaping corrections which result from the mean-field part of the e-e interaction. Our contention is that within our Kohn-Sham 
Kubo-Greenwood prescription: 
\begin{equation}
\label{eqsigmaDFT}
\sigma_{\rm DFT}= S_{\sigma}\sigma_{ei} = \sigma,
\end{equation}
\begin{equation}
\label{eqkappaDFT}
\kappa_{\rm DFT}= S_{\kappa}\kappa_{ei} = \frac{\kappa}{1 -{\kappa}/{\kappa^{\rm OCP}_{ee}}},
\end{equation}
where $\sigma$ and $\kappa$ are the true conductivities for the hydrogen plasma, i.e., as predicted by quantum-LB if we assume it to be perfectly 
valid in the conditions of interest.

Before we motivate Eqs.~\ref{eqSsigma}, \ref{eqSkappa}, \ref{eqsigmaDFT}, and \ref{eqkappaDFT} further with direct numerical comparisons, we must clarify 
what we mean by $\sigma_{ei}$ and $\kappa_{ei}$ here. Consider the QLB calculations of $\sigma$ and $\kappa$ for hydrogen. As mentioned above, the 
kinetic equation for the electron distribution function has two collision terms, $C_{ei}$ and $C_{ee}$, each of which involves the 2-component dielectric 
function, $\epsilon$. This dielectric function depends on all three fundamental interactions, $\phi_{ee}$, $\phi_{ii}$, $\phi_{ei}$ \cite{Ichimaru92}. The collision 
terms, $C_{ei}$ and $C_{ee}$, involve the screened interactions $(\phi_{ei}/\epsilon)$ and $(\phi_{ee}/\epsilon)$, respectively. The conductivities $\sigma$ 
and $\kappa$ are obtained by including both collision terms, while $\sigma_{ei}$ and $\kappa_{ei}$ are obtained by including only $C_{ei}$. However, it is 
important to note that $\sigma_{ei}$ and $\kappa_{ei}$ still include the effects of the e-e interaction within the dielectric function which screens $\phi_{ei}$, 
causing it to be reduced relative to its bare value. This inclusion is crucial, and is taken into account in many theories less sophisticated than LB, such as 
Spitzer-H\"arm (embedded in their assumption, $b_{\rm max}=$ Debye screening length) \cite{Spitzer53}, and the various Ziman formula approaches in which 
the effective electron-ion scattering potential is taken to be $\phi_{ei}/\epsilon_{\rm electron}$ \cite{Ziman61,Sterne07,Rosznyai08,Ichimaru85}. It is also 
accounted for in the rather sophisticated quantum-Boltzmann approach of Refs.~\cite{Kuhlbrodt00,Reinholz15} where the fundamental interaction within their 
$C_{ei}$ is $\phi_{ei}$ statically screened by the electrons.

The construction of the various terms in Eqs.~\ref{eqSsigma}, \ref{eqSkappa}, \ref{eqsigmaDFT}, and \ref{eqkappaDFT} from LB is then straightforward: 
Quantum-LB calculations including both $C_{ee}$ and $C_{ei}$ produce $\sigma$ and $\kappa$; calculations including only $C_{ei}$ produce $\sigma_{ei}$ 
and $\kappa_{ei}$. Calculations in which we apply a multiplier to $C_{ei}$ to force it to zero give us the electron OCP results, $\sigma^{\rm OCP}_{ee}$ and 
$\kappa^{\rm OCP}_{ee}$. As discussed, we recover $\sigma^{\rm OCP}_{ee} \to \infty$ for all densities and temperatures, while we obtain finite values for 
$\kappa^{\rm OCP}_{ee}$ as expected \cite{Chapman70}.

The need for the reshaping correction factors, $S_{\sigma}$ and $S_{\kappa}$, is obviated within LB by first comparing $\sigma$ with $\sigma_{ei}$. 
Table \ref{tablesigma} shows our quantum-LB results for hydrogen along the $\rho=$ 40 g/cm$^{3}$ isochore. The ratio $\sigma/\sigma_{ei}\equiv S_\sigma$ 
varies between 0.64 and 0.72 within this temperature range. Even though the inclusion of e-e interactions does nothing to degrade the electrical current of an 
OCP, the inclusion of $C_{ee}$ here reduces $\sigma$ by an appreciable amount, and this occurs even as the contributions of the e-e interaction within the 
screening function, $\epsilon$, are left unchanged. This reduction is due to the tendency for the e-e interaction to make the electron distribution function more 
isotropic in velocity space, which is seen clearly when the full solution is obtained by expanding $f$ in polynomials \cite{Williams69,Morales89,Whitley15} 
using the standard Chapman-Enskog procedure \cite{Chapman70}. The fact that our Kohn-Sham DFT electrical conductivities agree quite well with the 
quantum-LB $\sigma$, and far worse for $\sigma_{ei}$ (see Fig.~\ref{figsigma}), indicates that this reshaping effect is within the purview of a self-consistent 
mean-field Hartree/Vlasov approach \cite{caveat_xc}. Our assertion appearing in Eq.~\ref{eqsigmaDFT} is therefore justified.

\begin{figure}[!h]
\begin{center}
\includegraphics[width=0.47 \textwidth]{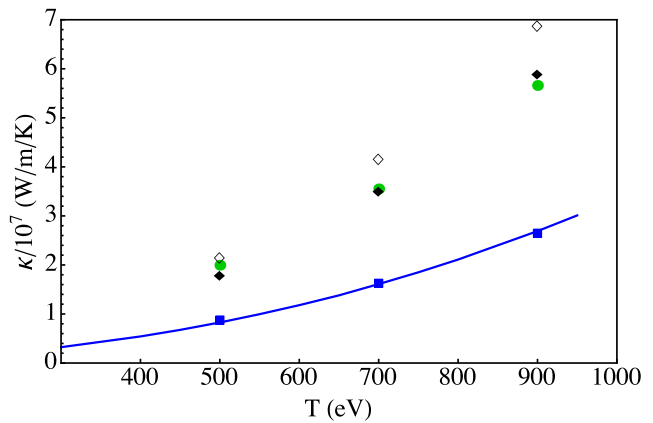}
\caption{(Color online) Thermal conductivity of hydrogen at $\rho = 40$~g/cm$^{3}$ as computed within a number of approximations.  QLB result 
for $\kappa$ (blue curve); $\kappa_{\rm DFT}$ extrapolated to an infinite number of single-particle states (solid green circles); $S_{\kappa}
\kappa_{ei}$ as determined from QLB (solid black diamonds); $S_\sigma\kappa_{ei}$ as determined from QLB (open diamonds); 
$\kappa_{\rm DFT}$ corrected with $\kappa_{ee}$ from QLB using Eq.~\ref{finalkappa} (solid blue squares).}
\label{figS}
\end{center}
\end{figure}

Turning to $\kappa$, if we now {\it assume} the relation of Eq.~\ref{eqSkappa}, our LB computations of $\kappa$, $\kappa_{ei}$, and 
$\kappa^{\rm OCP}_{ee}$ allow us to solve for $S_{\kappa}$. 
Fig.~\ref{figS} shows the product, $S_{\kappa}\cdot\kappa_{ei}$, vs. $T$ as the solid black diamonds. These are extremely close to our results for 
$\kappa_{\rm DFT}$, shown as the solid green circles. This justifies our supposition of Eq.~\ref{eqkappaDFT}, and points to a way to correct our 
Kohn-Sham DFT results for the thermal conductivity of hydrogen plasmas:
\begin{equation}
\label{finalkappa}
\frac{1}{\kappa}= \frac{1}{\kappa_{\rm DFT}} + \frac{1}{\kappa^{\rm OCP}_{\rm ee}}.
\end{equation}
The nearly coincident blue curve and solid blue squares at the bottom of Fig.~\ref{figS} shows the comparison of the $\kappa_{\rm QLB}$ with 
that of $\kappa_{\rm DFT}$ when corrected in this manner. 

In passing, we note that our quantum-LB results for hydrogen show that the reshaping correction factors for $\sigma$ and $\kappa$ are quite similar: 
$S_{\sigma} \sim S_{\kappa}$. This is illustrated by the relative closeness of $S_\sigma\kappa_{ei}$ (open diamonds) to $S_\kappa\kappa_{ei}$ (solid black 
diamonds) in Fig.~\ref{figS}.  Likewise, Table \ref{tableS} displays $\kappa$ as computed by Eq.~\ref{eqSkappa} but with $S_{\sigma}$ used instead of 
$S_{\kappa}$. Differences are less than $10\%$ and are decreasing as $T$ is increased. As shown in the Appendix, we have also used the Zubarev 
quantum-Boltzmann prescription of Refs.~\cite{Roepke88,Redmer97,Kuhlbrodt00} to affect the decompositions in Eqs.~\ref{eqSsigma} and \ref{eqSkappa}, 
and within this approach $S_{\sigma}$ and $S_{\kappa}$ are the same to within $5\%$ in the Spitzer limit for these conditions.  Alternatively we show that 
under the assumption $S_{\sigma} = S_{\kappa}$, the ansatz  
of Eq.~\ref{eqSkappa} is satisfied to within 3\% at the level of 4 moments, and within 2.2\% in the Spitzer limit.

We emphasize that we are able to assert the efficacy of the correction in Eq.~\ref{finalkappa} only because we are operating in a regime where we expect 
Lenard-Balescu to be accurate. One might then ask: Why use DFT at all, if LB is assumed to be better? The answer is that for stronger-coupling and/or for 
plasmas and conditions for which (unlike in the present cases) ionization is incomplete, DFT is sure to provide much benefit over LB, since LB as such is only 
able to describe weakly-coupled plasmas with no bound states, etc. Nevertheless, our primary aim in this work is to point out that {\it an uncorrected thermal 
conductivity from Kohn-Sham DFT \cite{Hansen11,Lambert11,Holst11,Hu14,Hu16} is very possibly incomplete in its description, and that some relation like 
that of Eq.~\ref{finalkappa} which accounts for the effects of explicit e-e collisions may be more appropriate}. It is also possible that for higher-Z plasmas, such 
as those studied in Ref.~\cite{Hu16}, the larger Z may cause the e-i interactions to outweigh the e-e interactions to the point where such effects are 
significantly less important \cite{Braginski58}. This will very likely depend on density and temperature, and additionally so because highly degenerate 
electrons will feel minimal effects from e-e scattering. More work must be done to further investigate these issues.

\begin{table}
\begin{center}
\begin{tabular}{| c | c | c | c |}
\hline
$k_{\rm B}T$ (eV)  & $\sigma$ (1/Ohm$\cdot$m) & $\sigma_{ei}$ (1/Ohm$\cdot$m) & $S_{\sigma}$ \\ \hline
500 & 1.31$\times 10^{8}$ & 1.82$\times 10^{8}$ & 0.72 \\ \hline
700 & 1.82$\times 10^{8}$ & 2.61$\times 10^{8}$ & 0.70 \\ \hline
900 & 2.37$\times 10^{8}$ & 3.45$\times 10^{8}$ & 0.69 \\ \hline
1000 & 2.65$\times 10^{8}$ & 3.89$\times 10^{8}$ & 0.68 \\ \hline
2000 & 5.76$\times 10^{8}$ & 8.76$\times 10^{8}$ & 0.66 \\ \hline
3000 & 9.31$\times 10^{8}$ & 1.44$\times 10^{9}$ & 0.65 \\ \hline
4000 & 1.32$\times 10^{9}$ & 2.05$\times 10^{9}$ & 0.64 \\ \hline
\end{tabular}
\caption{$\sigma$ and $\sigma_{ei}$ as determined from quantum Lenard-Balescu for hydrogen at $\rho= $ 40~g/cm$^{3}$; 
$S_{\sigma} = \sigma/\sigma_{ei}$. See text for details.}
\label{tablesigma}
\end{center}
\end{table}

\begin{table}
\begin{center}
\begin{tabular}{| c | c | c | c |}
\hline
$k_{\rm B}T$ (eV)  & $\kappa$ (W/m/K)& $  \kappa_{S_{\sigma}}$(W/m/K) & $\%$ difference\\ \hline
500 &  8.27$\times 10^{6}$ & 8.98$\times 10^{6}$ & 8.5 \\ \hline
700 &  1.61$\times 10^{7}$ & 1.74$\times 10^{7}$ & 8.1 \\ \hline
900 &  2.69$\times 10^{7}$ & 2.89$\times 10^{7}$ & 7.4 \\ \hline
1000 &  3.35$\times 10^{7}$ & 3.59$\times 10^{7}$ & 7.2 \\ \hline
2000 &  1.47$\times 10^{8}$ & 1.56$\times 10^{8}$ & 6.1 \\ \hline
3000 &  3.60$\times 10^{8}$ & 3.78$\times 10^{8}$ & 5.0 \\ \hline
4000 &  6.84$\times 10^{8}$ & 7.17$\times 10^{8}$ & 4.8 \\ \hline
\end{tabular}
\caption{Various quantities for hydrogen as computed with quantum Lenard-Balescu at $\rho = 40$~g/cm$^{3}$ :   
$\kappa$ ; $\kappa_{S_{\sigma}}$ ($\kappa$ as computed from
Eq.~\ref{eqSkappa}, but with $S_{\sigma}$ instead of $S_{\kappa}$); the percent difference between $\kappa_{S_{\sigma}}$ 
and $\kappa$. See text for details. }
\label{tableS}
\end{center}
\end{table}

\section{Conclusion}
We have presented an investigation of the electrical and thermal conductivities of hydrogen plasmas for $\rho= 40$~g/cm$^{3}$ and $T$ between 
500 eV and 900 eV using Kohn-Sham DFT together with a Kubo-Greenwood response framework to compute the relevant current-current 
correlation functions. In order to obtain converged results especially for the thermal conductivity, it was necessary to conduct a detailed 
extrapolation of transition dipole matrix elements to arrive at the results corresponding to an infinite number of high-lying Kohn-Sham states. The 
resulting electrical conductivities are in excellent agreement with the predictions of quantum Lenard-Balescu theory, while the thermal conductivities 
are roughly a factor of two larger than the Lenard-Balescu values. By conducting separate Lenard-Balescu studies in which electron-ion and 
electron-electron collision terms are independently switched off, we argue that the discrepancy in the thermal conductivity results from the neglect 
of explicit two-body electron-electron collisions in the (effectively mean-field) DFT prescription. In contrast, the electrical conductivity is well-
predicted by the DFT, suggesting that the well-known effect of the reshaping of the electron distribution function for that quantity is appropriately 
handled at the Hartree or Vlasov level. 

We propose the following correction to the thermal conductivity as predicted by Kohn-Sham DFT, at least for hydrogen plasmas: 
$1/\kappa = 1/\kappa_{\rm DFT} + 1/\kappa^{\rm OCP}_{ee}$, where $\kappa^{\rm OCP}_{ee}$ is the thermal conductivity of the electron 
one-component plasma at the same $(\rho,T)$. It remains to be seen if such a correction is sensible for plasmas other than hydrogen. In particular, 
it is not clear as to what should replace $1/\kappa^{\rm OCP}_{ee}$ for matter in which the ``free" electron density is less approximately 
represented by an electron OCP. Recent work on the electrical conductivity of warm, dense iron \cite{Pourovskii16} has used a correction supplied 
by Dynamical Mean Field Theory, and there are other works in which corrections to mean-field electronic structure approaches have been proposed 
along similar lines \cite{Domps98}. More fundamentally, it is likely of great interest to know if a more consistent formulation within the rubric of 
Time-Dependent DFT \cite{Gross85} and/or Current-DFT \cite{current} might admit a framework in which explicit electron-electron scattering can 
appear naturally in linear transport. These important questions we leave for future studies.

\bigskip

\section*{Acknowledgments}  The authors gratefully thank John~Castor, Frank Graziani, Gerd R{\"o}pke, Heidi Reinholz, and 
Martin French for helpful discussions.  RR thanks the Deutsche Forschungsgemeinschaft (DFG) for support within the SFB 652.
Sandia National Laboratories is a multi-program laboratory managed and operated by Sandia Corporation, a wholly owned subsidiary of Lockheed 
Martin Corporation, for the U.S. Department of Energy's National Nuclear Security Administration under contract DE-AC04-94AL85000. Portions of 
this work were performed under the auspices of the U.S. Department of Energy by Lawrence Livermore National Laboratory under contract 
DE-AC52-07NA27344, and were funded by the Laboratory Directed Research and Development Program at LLNL under tracking code 
No. 12-SI-005 as part of the Cimarron Project.

\appendix*
\section{}

The electrical and thermal conductivity are well known in the classical non-degenerate limit using kinetic theory, 
see e.g.,\ Spitzer~\cite{Spitzer53} and Chapman-Enskog~\cite{Chapman70}, or linear response theory as outlined by 
Zubarev which will be employed here; for details, see~\cite{Zubarev, Roepke88, Redmer97, Kuhlbrodt00,Reinholz15,RRN95}.
In the quantum-Boltzmann approach of Zubarev, which contains both the Ziman theory and the Spitzer theory as limiting cases,
the conductivities are determined in linear response theory to arbitrary order in generalized momenta, while 
permitting arbitrary electron degeneracy and strong scattering.  For the analysis here, the collision integrals
are regularized through the assumption of statically screened Coulomb potentials and the 
corresponding introduction of a Coulomb logarithm.  The electrons are assumed to be non-degenerate.

In particular, the influence of electron-electron collisions can be studied in order to validate the ansatz~(\ref{eqSkappa}) 
and the relation for the prefactors $S_\sigma \sim S_\kappa$. We start from the definition of the 
conductivities,
\begin{equation}
\label{eq:cond-def}
 \sigma = e^2 L_{11} \;,\; \kappa = \frac{1}{T} \left( 
 L_{22} - \frac{L_{12}L_{21}}{L_{11}} \right) ,
\end{equation} 
where the Onsager coefficients $L_{ik}$ are defined as
\begin{equation}
\label{eq:Lik}
L_{ik} = - \frac{h^{(i+k-2)}}{\Omega_0 \mid d \mid} \;
\begin{array}{|cc|}
  0 & \frac{k-1}{\beta h} \hat{N}_1 - \hat{N}_0 \\
  \frac{i-1}{\beta h} N_1 - N_0 & (D) 
\end{array} \;;
\end{equation}
$h$ denotes the enthalpy per particle, $\beta=1/k_BT$, and $\Omega_0$ is the system volume. 
The vectors $\hat{N}_n$, $N_n$ and the matrix $(D)$ in Eq.~(\ref{eq:Lik}) contain correlation functions 
which can be calculated for arbitrary densities and temperatures, see~\cite{Reinholz15,RRN95}. Using 
a finite set of $P$ moments to calculate the conductivities (i.e.,\ the nonequilibrium distribution 
function) we have 
\begin{eqnarray}
\label{eq:KFs}
 \hat{N}_m &=& ( \hat{N}_{m0}, \hat{N}_{m1}, \ldots , \hat{N}_{mP} ) 
         \;,\nonumber\\
 N_m &=& \left( \begin{array}{c} 
       N_{om} \\ N_{1m} \\ \vdots \\ N_{Pm} 
       \end{array} \right)\!,
 (D) = \left( \begin{array}{ccc} 
       D_{00} & \ldots & D_{0P} \\
       \vdots & \ddots & \vdots \\ 
       D_{P0} & \ldots & D_{PP}
       \end{array} \right)\!.
\end{eqnarray}
Generalized moments of the electron system are used to calculate the correlation functions,
\begin{equation}
\label{eq:Pn}
 {\bf P}_n = \sum_{\bf k} \hbar {\bf k} [\beta E_e(k)]^n a_e^{\dagger}(k) a_e(k) ,
\end{equation}
and the time derivatives $\dot{\bf P}_n = \frac{i}{\hbar}[H_s,{\bf P}_n]$. 
$H_s$ is the Hamilton operator of the system, the kinetic energy of the electrons is 
$E_e(k) = \hbar^2 k^2/(2m_e)$, and $a_e^{\dagger}(k)$ and $a_e(k)$ are creation and annihilation
operators for electronic states $k$, respectively. The correlation functions are given as Kubo 
scalar products and its Laplace transforms:
\begin{eqnarray}
\label{eq:corrfunc}
  N_{nm} &=& \frac{1}{m_e}({\bf P}_n,{\bf P}_m) ,\nonumber\\
  \hat{N}_{nm} &=& N_{nm} + \frac{1}{m_e} \langle
  {\bf P}_n(\varepsilon);\dot{\bf P}_m \rangle ,\\
  D_{nm} &=& \langle \dot{\bf P}_n(\varepsilon);\dot{\bf P}_m \rangle .\nonumber
\end{eqnarray}
In the nondegenerate limit, the terms $\langle {\bf P}_n(\varepsilon);\dot{\bf P}_m \rangle$ can be 
neglected since they are related to the Debye-Onsager relaxation effect and we have $N_{nm}=\hat{N}_{nm}$.
According to the Hamilton operator $H_s=T+V_{ei}+V_{ee}$, the force-force correlation functions 
$D_{nm}$ in Eq.~(\ref{eq:corrfunc}) can be separated with respect to electron-electron and electron-ion
scattering, i.e.\ $D_{nm} = D_{nm}^{\rm ee} + D_{nm}^{\rm ei}$, for which analytical expressions can be
given for hydrogen plasma ($N_i=N_e$) in the nondegenerate limit, see \cite{Kuhlbrodt00,Reinholz15,Reinholz89}:
\begin{eqnarray}
\label{eq:class-KF}
 N_{nm} &=& N_e \frac{\Gamma(n+m+5/2)}{\Gamma(5/2)} ,\\
 D_{nm} &=& d \, \left\{ \left( \frac{n+m}{2} \right)! + c_{nm}^{ee} \sqrt{2} \right\} ,\\
 d &=& \frac{4}{3} \sqrt{2\pi} \frac{e^4}{(4\pi\varepsilon_0)^2} 
       \frac{\sqrt{m_e}}{(k_BT)^{3/2}} n_e N_i \Phi(\Lambda) ,
\end{eqnarray}
with the Coulomb logarithm $\Phi(\Lambda)$. The weighting factors for the e-e correlation functions 
are given by $c_{0m}^{ee}=c_{m0}^{ee}=0$, $c_{11}^{ee}=1$, $c_{12}^{ee}=c_{21}^{ee}=11/2$, $c_{22}^{ee}=157/4$, $\ldots$
The conductivities can be represented as
\begin{eqnarray}
 \sigma &=& f \sigma^\ast ,\; 
 \sigma^\ast = \frac{(4\pi\varepsilon_0)^2 (k_BT)^{3/2}}{\sqrt{m_e} e^2 \Phi(\Lambda)} ,\label{eq:Spitzer-sigma}\\
 \kappa &=& L \left( \frac{k_B}{e} \right)^2 T \sigma ,\label{eq:Spitzer-lambda}
\end{eqnarray}
where $L$ is a Lorenz number.
The Spitzer theory~\cite{Spitzer53} gives the correct values in this limit with 
$f_{\rm Sp}^{ei+ee}=0.5908$ and $L_{\rm Sp}^{ei+ee}=1.6220$ if e-i and e-e interactions are considered. 
In the case of a Lorentz gas, i.e.,\ neglecting e-e scattering, we get the values 
$f_{\rm Sp}^{ei}=1.0159$ and $L_{\rm Sp}^{ei}=4.0$. The prefactors for solutions up to 4th order within 
the Zubarev approach are given in Table~\ref{tab-Sp}. They demonstrate a rapid convergence against the 
Spitzer values for the Lorentz gas and the fully interacting electron-ion system; see~\cite{Reinholz89,Adams07}.

\begin{table}[ht] 
\caption{Prefactors for the electrical ($f$) and thermal conductivity ($L$) according to Eqs.~(\ref{eq:Spitzer-sigma}) 
 and (\ref{eq:Spitzer-lambda}) in the nondegenerate limit. The Zubarev approach using an increasing number of moments 
 $P_n$ (\ref{eq:Pn}) is compared with the correct Spitzer values, see~\cite{Adams07,Reinholz89}. Furthermore, the 
 Lorenz number $\ell^{ee}$ of an electron OCP defined by Eq.~(\ref{eq:ee-OCP}) is given. Finally, the thermal conductivity 
 according to the ansatz (\ref{eqSkappa}) can be expressed by the Lorenz number $\ell^{ei+ee}$ defined in 
 Eqs.~(\ref{eqSkappaA}) - (\ref{eqSkappaB}).} 
 \label{tab-Sp}
\begin{center}
\begin{tabular}{lcccccc} 
\hline\hline
       & $f^{ei}$ & $f^{ei+ee}$ & $L^{ei}$ & $L^{ei+ee}$ & $\ell^{ee}$ & $\ell^{ei+ee}$ \\ \hline
Spitzer theory       & 1.0159 & 0.5908 & 4.0    & 1.6220 & - & - \\ \hline
$P_0$                & 0.2992 & 0.2992 & -      & -      & - & - \\
$P_0, P_1$            & 0.9724 & 0.5781 & 0.5971 & 0.6936 & 1.3223 & 0.4734 \\ 
$P_0, P_1, P_2$      & 1.0145 & 0.5834 & 3.6781 & 1.6215 & 1.6529 & 1.6004 \\ 
$P_0, P_1, P_2, P_3$ & 1.0157 & 0.5875 & 3.9876 & 1.6114 & 1.6716 & 1.6605 \\ \hline\hline
\end{tabular}
\end{center}
\end{table}

We now calculate the conductivities for an electron OCP model with only e-e interactions. 
This can be done with the Hamilton operator 
\begin{equation}
 H_s = T + \epsilon V_{ei} + V_{ee} 
\end{equation}
by taking the limit $\epsilon\to 0$ after calculating the $L_{ik}$. Otherwise, the Onsager coefficients are divergent 
($L_{11}$) or indefinite ($L_{12}$, $L_{22}$) in the nondegenerate limit. We have treated the electron OCP model by 
using up to four moments $P_n$. The result for the thermal conductivity can be represented as:
\begin{eqnarray}
\label{eq:ee-OCP}
 \kappa^{ee} &=& 
 \kappa^{\rm OCP}_{ee} = \left( \frac{k_B}{e} \right)^2 T \sigma^\ast \cdot \ell^{ee} .
\end{eqnarray} 
The values for the factor $\ell^{ee}$ are given in Table~\ref{tab-Sp}.
We now explore the ansatz (\ref{eqSkappa}) in the nondegenerate limit
\begin{eqnarray}
\label{eqSkappaA}
  \frac{1}{\kappa} &\stackrel{?}{=}& \frac{1}{S_{\kappa}\kappa_{ei}} + \frac{1}{\kappa^{\rm OCP}_{ee}} ,
 \end{eqnarray}
 within the approximation $S_{\kappa} = S_{\sigma} $.
We begin by writing 
  \begin{equation}
 S_{\kappa}\kappa_{ei} = L^{ei} S_{\kappa} \sigma_{ei} \left( \frac{k_B}{e} \right)^2 T  \approx L^{ei} S_{\sigma} \sigma_{ei} 
 \left( \frac{k_B}{e} \right)^2 T.
  \end{equation}
  
 Noting that 
$ \sigma^* = \sigma_{ei}/{f^{ei}} $ and $ S_{\sigma} = f^{ei + ee}/f^{ei}$
  we can rewrite the ansatz as 
  \begin{equation}
  \label{eqSkappaB}
   {\kappa}  = {\sigma\left( \frac{k_B}{e} \right)^2 T} \ell^{ei+ee}.  
 \end{equation}  
 where
 \begin{equation}
  \ell^{ei+ee}= \frac{L^{ei}\ell^{ee}}{\ell^{ee} + f^{ei+ee}L^{ei}}.
 \end{equation}
  
The factor $\ell^{ei+ee}$, to be compared to the exact value $L^{ei+ee}$, is listed in the last column of Table~\ref{tab-Sp}. 
We observe a fast convergence as before to 1.6605 at the level of 4 moments. 
The deviation from the correct value in 4th order (1.6114) is just 3.0\%,  
i.e., \ similar to the deviations of the numerical data from the quantum LB equation; see Figs.~\ref{figsigma} 
and \ref{figS}.  If instead we use the Spitzer values for $f^{ei+ee}$ and $L^{ei + ee}$ agreement with the
Spitzer Lorenz number is within 2.2\%.  
Note that the coincidence of the values for $L^{ei+ee}$ and $\ell^{ei+ee}$, i.e.,\ of the direct and sum 
of the inverse thermal conductivities representing electron-ion and electron-electron scattering 
contributions, is only obtained if we use the prefactor $S_\sigma=f^{ei+ee}/f^{ei}$, which contains the influence of 
both electron-ion and electron-electron scattering. 

Alternatively, we can take the ansatz (\ref{eqSkappa}) as an equality and solve for $S_\kappa/S_\sigma$
 \begin{equation}
 \frac{S_{\kappa}}{S_{\sigma}}= \frac{1}{L^{ei}}  \left(\frac{L^{ei+ee}\ell^{ee}} {\ell^{ee} - L^{ei+ee}f^{ei + ee}}\right) .
 \end{equation}
  At the level of 4 moments, we find $S_\kappa/S_\sigma = 0.9318$ and with the Spitzer values  for
  $f^{ei+ee}$, $L^{ei}$, and $L^{ei + ee}$, we obtain $S_\kappa/S_\sigma = 0.9502$.  Note that both of these results, 
  for $\ell^{ei+ee}$ or $S_\kappa/S_\sigma$, were obtained in a parallel approach to that taken in the main
  text within the QLB framework, and are completely general in the non-degenerate limit.


\begin{thebibliography}{100}
\bibitem{caveat_indeed}Indeed, it has even been shown that LB theory works quite well for considerably stronger plasma 
couplings when compared with classical MD, as demonstrated in the work of Ref.~\cite{Bernu88}.
\bibitem{caveat_WD}Our Lenard-Balescu treatment is very similar to that of Ref.~\cite{Williams69}, with the notable exception that they employ an additional collision term to account for hard scattering. We have established that this additional term is of no importance for the plasma conditions discussed here. We also note that our numerical procedures are slightly different from theirs for the terms which are common to the two approaches; but again, the discrepancies resulting from these differences are very minor. See Figs. 1 and 2 of Ref.~\cite{Whitley15} for comparisons along these lines.
\bibitem{degeneracy}By comparing the conductivities calculated with classical versus quantum collision integrals within the Zubarev formalism, we estimate this difference to be on the order of 20\% at 500 eV, and less than 10\% at 900 eV.
\bibitem{caveat_v}We use the notation $f(v)$ somewhat loosely here; $v$ refers to the velocity of the electron in the semiclassical sense. In the actual Kohn-Sham DFT calculation performed on a periodic array of supercells, $v$ is related to the wave vector, ${k}$.
\bibitem{caveat_xc}It should be noted that the exchange-correlation term within the Kohn-Sham Hamiltonian plays practically no role for hydrogen at these $(\rho,T)$ conditions.
\end{thebibliography}
\end{document}